\documentclass[aps,pre,showpacs,reprint,groupedaddress]{revtex4-1}
\usepackage{amsfonts}
\usepackage{amsmath}
\usepackage{graphicx}

\begin{document}

\title{Renewal stochastic processes with correlated events. Phase
transitions along time evolution.}
\author{Jorge Vel\'{a}zquez${}^{1}$}

\author{Alberto Robledo${}^{1,2,}$}
\email{robledo@fisica.unam.mx}
\affiliation{${}^{1}$Instituto de F\'{\i}sica, Universidad Nacional Aut\'{o}noma de
M\'{e}xico,
Apartado postal 20-364, M\'{e}xico 01000 D.F., Mexico \\ ${}^{2}$Departamento de
Matem\'aticas, Universidad Carlos III de Madrid, Spain  {\rm (}on sabbatical leave{\rm )}}
\date{\today}

\begin{abstract}
We consider renewal stochastic processes generated by non-independent events
from the perspective that their basic distribution and associated generating
functions obey the statistical-mechanical structure of systems with
interacting degrees of freedom. Based on this fact we look briefly into the
less known case of processes that display phase transitions along time. When
the density distribution $\psi _{n}(t)$ for the occurrence of the $n$-th
event at time $t$ is considered to be a partition function, of a
`microcanonical' type for $n$ `degrees of freedom' at fixed `energy' $t$,
one obtains a set of four partition functions of which that for the
generating function variable $z$ and Laplace transform variable $\epsilon $,
conjugate to $n$ and $t$, respectively, plays a central role. These
partition functions relate to each other in the customary way and in
accordance to the precepts of large deviations theory, while the entropy, or
Massieu potential, derived from $\psi _{n}(t)$ satisfies an Euler relation.
We illustrate this scheme first for an ordinary renewal process of events
generated by a simple exponential waiting time distribution $\psi (t)$. Then
we examine a process modelled after the so-called Hamiltonian Mean Field
(HMF) model that is representative of agents that perform a repeated task
with an associated outcome, such as an opinion poll. When a sequence of
(many) events takes place in a sufficiently short time the process exhibits
clustering of the outcome, but for larger times the process resembles that
of independent events. The two regimes are separated by a sharp transition,
technically of the second order. Finally we point out the existence of a
similar scheme for random walk processes.
\end{abstract}

\maketitle






\affiliation{Instituto de F\'{\i}sica, Universidad Nacional Aut\'{o}noma de M\'{e}xico,
Apartado postal 20-364, M\'{e}xico 01000 D.F., M\'{e}xico}





\section{Introduction}

A large class of stochastic processes are renewal processes \cite{renewal1}
\cite{renewal2}. This class of sequences are generally used to model
independent identically distributed (iid) occurrences. The renewal processes
are concerned with the times of substitution of components that are replaced
as soon as they break down. Here we recall \cite{recall1} \cite{recall2}
that this basic type of stochastic process possesses all the elements of a
statistical-mechanical system and therefore can be couched into this
language and benefit from well-established methods and applications
developed for the study of systems with many degrees of freedom. The common
iid process maps into the non-interacting case, but the most important
potential application of this equivalence is to the generalization of
renewal processes to correlated events, where the large body of knowledge
accumulated in the study of (short or long range) interacting particle or
spin systems can find interesting guidelines or clear-cut analogies for
renewal processes. One particular property that we present here is the
occurrence of phase transitions along time evolution.

The layout of the article is as follows. We start in Section 2 with a
concise description of a renewal process that involves the transformation
into Laplace space of the relevant probability density functions and the use
of generating functions \cite{montroll1}. In Section 3 we make explicit the
statistical-mechanical ensemble structure, with only two pairs of conjugate
variables, of the renewal process and illustrate the form that the partition
functions take for the simple case of an exponentially-decaying waiting-time
distribution density. In Section 4 we apply the saddle-point approximation
in the evaluation of the partition functions and show that the required
Legendre transform structure, the associated equations of state, and the
Euler relation \cite{callen1} are present in the formalism for the renewal
process. In Section 5 we consider a specific example of a renewal process
with correlated events that exhibits a phase transition when the time
variable increases. The renewal process is representative, for instance, of
an opinion poll, and is constructed to be equivalent to the
statistical-mechanical Hamiltonian Mean Field (HMF) model of interacting
particles \cite{ruffo1, ruffo2, ruffo3}. Finally in Section 6 we summarize
and discuss our results.

\section{Basics of renewal processes}

Technically, an \emph{ordinary} renewal process is a sequence of partial
sums of iid positive random variables. This process may be thought of as a
sequence of points in time when the lifetimes of some objects of the same
type ends and they are replaced by new ones. The renewal process counts the
number of renewals in the interval $[0,t)$, hence such a renewal counting
process is a random piecewise constant function. A convenient analytical
procedure to determine the properties of this kind of process is that of
Montroll \cite{montroll1}. It is resumed as follows: Let $\psi (t)$ be the
(normalized) waiting time distribution density for a single event and $\psi
_{n}(t)$ the distribution density for the occurrence of the $n$-th event at
time $t$. For iid events these densities are linked via

\begin{equation}
\psi _{n}(t)=\int\limits_{0}^{t}dt^{\prime }\ \psi (t-t^{\prime })\ \psi
_{n-1}(t^{\prime }),\ n>1,  \label{n-thpsi1}
\end{equation}%
or in Laplace space by%
\begin{equation}
\widehat{\psi }_{n}(\epsilon )=\left[ \widehat{\psi }(\epsilon )\right] ^{n},
\label{n-thpsilaplace1}
\end{equation}
where%
\begin{equation}
\widehat{\psi }(\epsilon )=\int\limits_{0}^{\infty }dt\ \exp (-\epsilon t)\
\psi (t)
\end{equation}
and
\begin{equation}
\widehat{\psi }_{n}(\epsilon )=\int\limits_{0}^{\infty
}dt\ \exp (-\epsilon t)\ \psi _{n}(t).  \label{laplacedefs1}
\end{equation}%
We shall consider throughout this paper time variables to be dimensionless.
A generating function for the $\psi_{n}(t)$ is defined via the $z$-transform%
\begin{equation}
\psi (t;z)\equiv \sum_{n=1}^{\infty }\psi _{n}(t)\ z^{n},
\label{generating1}
\end{equation}%
so that
\begin{equation}
\widehat{\psi }(\epsilon ;z)\equiv \int\limits_{0}^{\infty }dt\ \exp
(-\epsilon t)\ \psi (t;z)=\sum_{n=1}^{\infty }\widehat{\psi }_{n}(\epsilon
)\ z^{n}.  \label{laplacedef2}
\end{equation}%
Use of Eq. (\ref{n-thpsilaplace1}) above turns $\widehat{\psi }(\epsilon ;z)$
into a geometric series that when convergent becomes%
\begin{equation}
\widehat{\psi }(\epsilon ;z)=\left[ \widehat{\psi }(\epsilon )\ z\right] /%
\left[ 1-\widehat{\psi }(\epsilon )\ z\right].  \label{psiepsilonzetaiid1}
\end{equation}%
The functions $\psi _{n}(t)$ and $\psi (t;z)$ are recovered from $\widehat{%
\psi }_{n}(\epsilon )$ and $\widehat{\psi }(\epsilon ;z)$, respectively, via
inverse Laplace and inverse $z$ transforms. The average time between events,
or period, $T$ is given by the first moment of $\psi (t)$,%
\begin{equation}
T\equiv \int\limits_{0}^{\infty }dt\ t\ \psi (t)=\left. -\frac{d}{d\epsilon }%
\ln \widehat{\psi }(\epsilon )\right\vert _{\epsilon =0}\ <\infty ,
\label{period1}
\end{equation}%
whereas the average number of events $\left\langle n(t)\right\rangle $ of a
renewal sequence when the last event occurs at time $t$ \cite{note1} is%
\begin{equation}
\left\langle n(t)\right\rangle \equiv \frac{\sum_{n=1}^{\infty }n\ \psi
_{n}(t)}{\psi (t;1)}=\left. z\frac{d}{dz}\ln \psi (t;z)\right\vert _{z=1}\ .
\label{averagenumberevents1}
\end{equation}%
Therefore, if the most common calculation aim is to determine $\psi _{n}(t)$
or $\left\langle n(t)\right\rangle $ for any given waiting time $\psi (t)$
distribution, use of $\widehat{\psi }(\epsilon )$ in Eqs. (\ref%
{n-thpsilaplace1}) and (\ref{psiepsilonzetaiid1}) followed by inverse
transformation is an expedient method.

Here we recapture \cite{recall1} \cite{recall2} a precise interpretation of
the above expressions while calling attention that it is not restricted to
iid processes. This is that the functions $\psi _{n}(t)$, $\widehat{\psi }%
_{n}(\epsilon )$, $\psi (t;z)$, and $\widehat{\psi }(\epsilon ;z)$ can be
seen to be partition functions associated to an equilibrium
statistical-mechanical system of $n$ degrees of freedom arranged in
configurations with energy measured by a time $t$. Below we detail that in
the large $t$ and $n$ limits these functions can be evaluated via the
saddle-point approximation and that this central statistical-mechanical
property leads to equations of state and entropies or free energies related
via Legendre transforms, where the variables $\epsilon $ and $\mu \equiv \ln
z$ appear to be conjugate to the variables $t$ and $n$, respectively.

\section{Statistical ensembles for renewal processes}

We observe that Eq. (\ref{laplacedef2}), $\widehat{\psi }(\epsilon
;z)=\sum_{n=1}^{\infty }\widehat{\psi }_{n}(\epsilon )\ z^{n}$, has the form
of the expression for the grand canonical partition function of a thermal
system if we were to consider that the number of events $n$ represents the
number of particles or degrees of freedom, $\epsilon $ the inverse
temperature, $z$ the activity, and therefore $\widehat{\psi }_{n}(\epsilon )$
plays the role of the canonical partition function. Having considered that $%
\widehat{\psi }_{n}(\epsilon )$, the Laplace transform of $\psi _{n}(t)$ in
Eq. (\ref{laplacedefs1}), plays this role, the formal analogy can be
extended by identification of $\psi _{n}(t)$ as the microcanonical partition
function where $t$ is the energy. Further, the generating function $\psi
(t;z)$ would then be seen as the partition function corresponding to an
ensemble of fixed energy $t$ and activity $z$. (For iid random variables $%
\widehat{\psi }_{n}(\epsilon )$ is given by Eq. (\ref{n-thpsilaplace1}) and
the corresponding thermal system is made of identical non-interacting
degrees of freedom with $\widehat{\psi }(\epsilon )$ the canonical partition
function per degree of freedom).

The scope of this analogy can be further assessed by defining the following
entropies or Massieu potentials \cite{callen1},
\begin{equation}
\begin{array}{cc}
\rule[-3ex]{0pt}{0pt} S_{\epsilon ,\mu }\equiv \ln \widehat{\psi }(\epsilon ;z), & S_{\epsilon
,n}\equiv \ln \widehat{\psi }_{n}(\epsilon ), \\
 S_{t,\mu }\equiv \ln \psi
(t;z), & S_{t,n}\equiv \ln \psi _{n}(t), 
\end{array}
 \label{entropies1}
\end{equation}%
where we have introduced the `chemical potential' $\mu \equiv \ln z$. (These
quantities may be negative since the arguments of the logarithms may be less
than unity. Notice that these arguments are probability densities or their
Laplace and/or $z$-transforms, while in ordinary statistical mechanics the
arguments are configuration numbers or their transforms). If for large $n$ a
thermodynamic limit or a large deviations property \cite{touchette1} arises,
then these potential functions would be related via Legendre transforms
involving the pairs of conjugate variables $(n,\mu )$ and $(t,\epsilon )$
and mediated via the corresponding equations of state. An Euler relation of
the type%
\begin{equation}
S_{t,n}=t\epsilon -n\mu  \label{euler1}
\end{equation}%
would hold, and attention should be paid in the evaluation of $S_{\epsilon
,\mu }=\ln \widehat{\psi }(\epsilon ;z)$ as a cursory inspection of repeated
Legendre transforms would imply $S_{\epsilon ,\mu }=S_{t,n}-t\epsilon +n\mu
=0$. Below we show that Eq. (\ref{euler1}) holds with a nonzero $S_{\epsilon
,\mu }$.

To help us examine the validity of this formal structure in the following
section we determine the above partition functions for the particular iid
random variable case of an exponential waiting-time density $\psi (t)=b\exp
(-bt)$. One obtains%
\begin{equation}
\widehat{\psi }(\epsilon ;z)=bz\left( b+\epsilon -bz\right) ^{-1},
\label{psiepsilonzeta1}
\end{equation}%
\begin{equation}
\widehat{\psi }_{n}(\epsilon )=b^{n}\left( b+\epsilon \right) ^{-n},
\label{psiepsilonene1}
\end{equation}%
\begin{equation}
\psi (t;z)=bz\exp (-bt+btz)  \label{psitezeta1}
\end{equation}%
and%
\begin{equation}
\psi _{n}(t)=\frac{\left( bt\right) ^{n}}{n!}\exp (-bt),  \label{psiteene1}
\end{equation}%
where we recognize in the last equation the distribution density of a
Poisson process.

\section{Analogy with statistical mechanics}

The asymptotic solution of $\widehat{\psi }_{n}(\epsilon )$ for $n>>1$ can
be found by use of the steepest-descent approximation of the inverse $z$%
-transform of $\widehat{\psi }(\epsilon ;z)$,
\begin{eqnarray}
\widehat{\psi }_{n}(\epsilon )=\frac{1}{2\pi i}\oint dz&&\ \exp \left[
(n-1)\left(-\ln z+ \right.\right.  \nonumber\\
&& \left.\left. (n-1)^{-1}\ln \widehat{\psi }(\epsilon ;z)\right) \right],
  \label{epsilonene1}
\end{eqnarray}
One obtains%
\begin{equation}
\ln \widehat{\psi }_{n}(\epsilon )\simeq -(n-1)\ln z_{0}+\ln \widehat{\psi }%
(\epsilon ;z_{0}),  \label{epsilonene2}
\end{equation}%
where $z_{0}$ can be eliminated in favor of $n$ via the steepest-descent
condition%
\begin{equation}
n-1=\left. z\frac{d}{dz}\ln \widehat{\psi }(\epsilon ;z)\right\vert
_{z=z_{0}}.  \label{epsilonene3}
\end{equation}%
It is then possible to write%
\begin{equation}
\widehat{\psi }_{n}(\epsilon )=\exp S_{\epsilon ,n},  \label{epsilonene4}
\end{equation}%
where $S_{\epsilon ,n}=$ $\ln \widehat{\psi }_{n}(\epsilon )$ is the
Legendre transform $S_{\epsilon ,n}=$ $-(n-1)\mu _{0}+S_{\epsilon ,\mu _{0}}$
of $S_{\epsilon ,\mu _{0}}=\ln \widehat{\psi }(\epsilon ;z_{0})$. For the
exponential waiting-time density $\psi (t)=b\exp (-bt)$ this transformation
leads to the equation of state%
\begin{equation}
n-1=1+bz_{0}(b+\epsilon -bz_{0})^{-1},  \label{epsilonene5}
\end{equation}%
and to the Massieu potential%
\begin{equation}
S_{\epsilon ,n}=\ln \left[ \frac{b^{n-1}}{(b+\epsilon )^{n-1}}\frac{%
(n-1)^{n-1}}{(n-2)^{n-2}}\right] .  \label{epsinonene6}
\end{equation}%
In the limit $n\rightarrow \infty $ Eq. (\ref{psiepsilonene1}) is recovered.

Similarly, the asymptotic solution of $\psi _{n}(t)$ for $n>>1$ is obtained
with the use of the steepest-descent approximation of the inverse Laplace
transform of $\widehat{\psi }_{n}(\epsilon )$,
\begin{equation}
\psi _{n}(t)=\frac{1}{2\pi i}\int\limits_{c-i\infty }^{c+i\infty }d\epsilon
\ \exp \left[ n\left( \epsilon \tau +\ln \widehat{\psi }(\epsilon )\right) %
\right] ,  \label{teene1}
\end{equation}%
where $\tau \equiv t/n$. One obtains%
\begin{equation}
n^{-1}\ln \psi _{n}(n\tau )\simeq \epsilon _{0}\tau +\ln \widehat{\psi }%
(\epsilon _{0}),\ n>>1,  \label{teene2}
\end{equation}%
where $\epsilon _{0}$ can be eliminated in favor of $\tau $ via the
steepest-descent condition%
\begin{equation}
\tau =\left. -\frac{d}{d\epsilon }\ln \widehat{\psi }(\epsilon )\right\vert
_{\epsilon =\epsilon _{0}}.  \label{teene3}
\end{equation}%
In taking the limit $n\rightarrow \infty $ also $t\rightarrow \infty $ but $%
\tau $ is kept finite. We can therefore write%
\begin{equation}
\psi _{n}(t)=\exp S_{t,n},  \label{teene4}
\end{equation}%
where $S_{t,n}=$ $\ln \psi _{n}(t)$ is the Legendre transform $S_{t,n}=$ $%
-t\epsilon _{0}+S_{\epsilon _{0},n}$ of $S_{\epsilon _{0},n}=\ln \widehat{%
\psi }_{n}(\epsilon _{0})$. For exponential waiting times $\psi (t)=b\exp
(-bt)$ this transformation leads to the equation of state%
\begin{equation}
t=n(b+\epsilon _{0})^{-1},  \label{teene5}
\end{equation}%
and to the Massieu potential%
\begin{equation}
S_{t,n}=\ln \left[ \left( btn^{-1}\right) ^{n}\exp (n)\exp (-bt)\right] .
\label{teene6}
\end{equation}%
Therefore Eq. (\ref{psiteene1}) is recovered in the limit $n\rightarrow
\infty $ (when we notice that the Stirling approximation of the factorial is
part of Eq. (\ref{teene6})).

Lastly, following an analogous procedure the asymptotic form for $\psi (t;z)$
for $n>>1$ is given by%
\begin{equation}
\psi (t;z)=\exp S_{t,\mu },  \label{tezeta1}
\end{equation}%
where the Massieu potential $S_{t,\mu }$ for $\psi (t)=b\exp (-bt)$,%
\begin{equation}
S_{t,\mu }=\ln \left[ btz\exp \left( 1+bt+btz\right) \right] ,
\label{tezeta2}
\end{equation}%
is obtained as the Legendre transform $S_{t,\mu }=$ $t\epsilon
_{0}+S_{\epsilon _{0},z}$ of $S_{\epsilon _{0},\mu }=\ln \widehat{\psi }%
(\epsilon _{0};z)$ with%
\begin{equation}
t=\left( b+\epsilon _{0}-bz\right) ^{-1}.  \label{tezeta3}
\end{equation}%
Eq. (\ref{tezeta2}) is consistent with Eq. (\ref{psitezeta1}) in the limit $%
n\rightarrow \infty $.

To make explicit the observance of the Euler relation Eq. (\ref{euler1}) we
note that the inverse Legendre transform that yields $S_{\epsilon ,\mu }$
from $S_{\epsilon ,n_{0}}$, $S_{\epsilon ,\mu }=$ $(n_{0}-1)\mu +S_{\epsilon
,n_{0}}$, requires%
\begin{equation}
\mu =-\left. \frac{d}{d(n-1)}n\ln \ \widehat{\psi }(\epsilon )\right\vert
_{n=n_{0}}=-\ln \widehat{\psi }(\epsilon ),  \label{mucero1}
\end{equation}%
so that $S_{\epsilon ,n_{0}}=$ $n_{0}\ln \widehat{\psi }(\epsilon
)=-n_{0}\mu $ and $S_{\epsilon ,\mu }=$ $(n_{0}-1)\mu +S_{\epsilon
,n_{0}}=-\mu $. This leads to%
\begin{equation}
S_{t_{0},n_{0}}=t_{0}\epsilon -(n_{0}-1)\mu +S_{\epsilon ,\mu
}=t_{0}\epsilon -n_{0}\mu .  \label{euler2}
\end{equation}%
We note that the existence of the Euler relation for a system\ with only two
pairs of conjugate variables does not imply the vanishing of the
thermodynamic potential, $S_{\epsilon ,\mu }$, associated to two consecutive
Legendre transforms of the basic potential, $S_{t,n}$, a homogeneous
function of order one in both variables $t$ and $n$ \cite{callen1}. Notably,
the partition function $\widehat{\psi }(\epsilon ;z)$ associated to the
variables $\epsilon $ and $\mu $ remains a fundamental and most useful
quantity for the description of the renewal process.

\section{An example of a renewal process with correlated events}

As an illustration of how developments in the statistical mechanics of
interacting particle or spin systems may have meaningful translations to
renewal processes we present here features of a renewal process with
correlated events. We take inspiration from the so-called Hamiltonian Mean
Field (HMF) model \cite{ruffo1}-\cite{ruffo3} to point out the occurrence of
phase transitions along time evolution.

Consider a sequence of events, each of which, besides taking place at a
given time $t$, assigns values to two scalar quantities $\tau $ and $\theta $%
, the first within the time interval $0\leq \tau \leq t$ taken by the event,
and the second within a fixed finite interval, say $0\leq \theta \leq 2\pi $%
. For instance, the process may represent an agent (or agents) that performs
a repeated task with outcome $(\tau ,\theta )$ that is not independent of
those for all the previous events. A string of such $n$ events is described
by the sequence of triplets $[(t_{1};\tau _{1},\theta _{1}),...,(t_{n};\tau
_{n},\theta _{n})]$. The two collections of values $(\tau _{1}$, $\tau _{2}$%
,..., $\tau _{n})$ and $(\theta _{1}$, $\theta _{2}$,..., $\theta _{n})$ are
used to construct two additional time variables, $T_{n}$, the
\textquotedblleft idle\textquotedblright\ time, and $W_{n}$, the
\textquotedblleft active\textquotedblright\ time, respectively, that
together comprise the total time taken by the sequence of $n$ events, i.e.%
\begin{equation}
t_{n}=T_{n}+W_{n}.  \label{totaltime}
\end{equation}%
The idle time $T_{n}$ is simply given by%
\begin{equation}
T_{n}=\sum_{i=1}^{n}\tau _{i},  \label{idle1}
\end{equation}%
whereas the active time $W_{n}$ measures the dispersion of the values $%
(\theta _{1}$, $\theta _{2}$,..., $\theta _{n})$ over the unit circle, being
large when these are spread out over $(0,2\pi )$ and small when they
concentrate around a given $\theta $. Although there are many options to
define $W_{n}$, for definiteness we chose it to be determined by%
\begin{equation}
W_{n}=\frac{1}{2n}\displaystyle\sum\limits_{i,j=1}^{n}\left[ 1-\cos
(\left\vert \theta _{i}-\theta _{j}\right\vert )\right] \leq t,
\label{effectiveness1}
\end{equation}%
where all pairs $(\theta _{i}$, $\theta _{j})$ are equally considered. (We
recall that time variables are considered dimensionless). Clearly, the
condition Eq. (\ref{totaltime}) imposes a restriction in the possible values
of the sequences $(\tau _{1}$, $\tau _{2}$,..., $\tau _{n})$ and $(\theta
_{1}$, $\theta _{2}$,..., $\theta _{n})$. As a more specific illustration of
this kind of process let us suppose there is an opinion poll organization
that sends an agent (or group of pollsters) to take a survey that consists
of a succession of $n$ completed questionnaires obtained in the time
interval $(0,t)$. Each respondent has a tag $\theta $ that quantifies a
characteristic of the population surveyed, such as age, race, home
environment, etc., and therefore $W_{n}$ reflects the degree of coverage
bias, in, for instance, the consideration of young voters, minorities, or
rural areas. The time $\tau $ associated to each respondent measures wasted
time in collecting opinions, since some people do not answer calls, or
refuse to answer the poll, or do not give candid answers, and consequently $%
T_{n}$ represents the extent of nonresponse bias. \

The probability density of occurrence of the $n$-th event at time $t$ with
outcome $(\theta ,\tau )$, $\psi _{n}(t;\tau ,\theta )$, is evaluated in
terms of the statistics of occurrence of the previous $n-1$ events. This is
best prescribed in terms of the Laplace transform of $\psi _{n}(t;\tau
,\theta )$ with respect to $t$, $\widehat{\psi }_{n}(\epsilon ;\tau ,\theta
) $. Specifically, the renewal process is given by%
\begin{equation}
\widehat{\psi }_{n}(\epsilon ;\tau ,\theta )\equiv \exp \left[ -\epsilon
\left\langle \delta t_{n}(\tau ,\theta )\right\rangle _{n-1}\right] ,
\label{nth-eventdensity1}
\end{equation}%
where the average $\left\langle ...\right\rangle _{n-1}$ is performed over
all possible values of $(\tau _{1},\theta _{1}),...,(\tau _{n-1},\theta
_{n-1})$, and
\begin{equation}
\delta t_{n}(\tau ,\theta )=T_{n}+W_{n}-T_{n-1}-W_{n-1}.
\label{insertiontime1}
\end{equation}

The analogy with the HMF model becomes evident when it is seen that Eq. (\ref%
{nth-eventdensity1}) corresponds to Widom's particle insertion formula when
applied to the thermal system (for vanishing chemical potential $\mu $) \cite%
{widom1} \cite{baldovin1}. The roles of the number of particles, their
positions (in the unit circle), inverse temperature, energy, kinetic energy,
and potential energy of the HMF model, are given, respectively, by $n$, $%
\theta _{i}$, $\epsilon $, $t_{n}$, $T_{n}$, and $W_{n}$. (With no loss of
generality a coupling constant in the potential energy term of the
ferromagnetic HMF model has been set equal to unity). In our notation, the
Helmholtz free energy of the HMF model in the limit $n\rightarrow \infty $,
obtained via the saddle-point approximation \cite{ruffo1}-\cite{ruffo3}, is%
\begin{equation}
-\epsilon \widehat{\psi }_{n}(\epsilon )=-\frac{1}{2}\ln \left( \frac{%
\epsilon }{2\pi }\right) -\frac{\epsilon }{2}+\max_{x}\left( -\frac{\epsilon
x^{2}}{2}+\ln 2\pi I_{0}(\epsilon x)\right)   \label{helmholtz1}
\end{equation}%
where the auxiliary variable $x$ satisfies%
\begin{equation}
x=\epsilon \frac{I_{1}(\epsilon x)}{I_{0}(\epsilon x)}  \label{auxilliary1}
\end{equation}%
and where $I_{i}(y)$ is the modified Bessel function of order $i$. As known
\cite{ruffo1}-\cite{ruffo3}, the HMF model exhibits two equilibrium phases,
indicated by the possible solutions of Eq. (\ref{auxilliary1}). When $%
\epsilon <\epsilon _{c}=2$ the variable $x$, identified as the model's
magnetization $M$, vanishes, but $x$ is nonzero for $\epsilon >\epsilon _{c}$%
, increasing gradually as $\epsilon $ increases and reaching unity as $%
\epsilon \rightarrow \infty $. These properties imply that the mean active
time $\left\langle W_{n}\right\rangle $, the average of $W_{n}$ over all
sequences $(\tau _{1},\theta _{1}),...,(\tau _{n},\theta _{n})$, is given by
$\left\langle W_{n}\right\rangle =(1-x^{2})n/2$, and that the relationship
between the time $t$ and the amplitude $\epsilon $ (the caloric equation for
the HMF model) reads%
\begin{equation}
t=\frac{n}{2\epsilon }+\left\langle W_{n}\right\rangle ,  \label{caloric1}
\end{equation}%
\cite{ruffo1}-\cite{ruffo3}. 
\begin{figure}[h]
\includegraphics{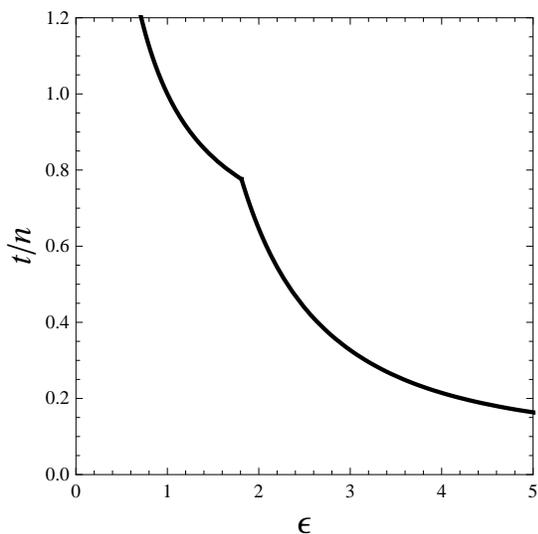}
\caption{Dependence of $t/n$, $t,n \gg 1$, on the Laplace variable $\protect%
\epsilon $ for the renewal process model of correlated events designed to be
analogous to the HMF model. The figure is equivalent to the caloric equation
of the HMF model and shows the two-phase behavior described in the text.}
\end{figure}
Thus, $\left\langle W_{n}\right\rangle $
displays a fixed maximum value $\left\langle W_{n}\right\rangle =n/2$ for $%
\epsilon <\epsilon _{c}=2$, whereas it decreases and approaches zero as $%
\epsilon \rightarrow \infty $. The two-phase behavior and its transition at $%
\epsilon _{c}$ is reflected by the $\theta $-dependence of $\widehat{\psi }%
_{n}(\epsilon ;\tau ,\theta )$ when $n\gg 1$. For small $\epsilon $ the
distribution is uniform in $\theta $, but when $\epsilon >\epsilon _{c}$ it
becomes peaked around a given (although arbitrary) value of $\theta =\phi $.
As we see below this feature is preserved in the distribution for the
original variable $t$, i.e. there is a critical time $t_{c}$ above which $%
\psi _{n}(t;\tau ,\theta )$ is uniform in $\theta $ and below which it is
peaked around a given $\theta =\phi $. The $\tau $-dependence of \ $%
\widehat{\psi }_{n}(\epsilon ;\tau ,\theta )$ 
\ has an exponential form (Gaussian if written for the `momentum' $p=\pm
\sqrt{2\tau }$ ) for all $\epsilon $. When $n\rightarrow \infty $ Eq. (\ref%
{nth-eventdensity1}) leads to \cite{ruffo3}%
\begin{equation}
\widehat{\psi }_{n}(\epsilon ;\tau ,\theta )=\sqrt{\frac{\epsilon }{2\pi }}%
\exp \left( -\epsilon \tau \right) \frac{1}{2\pi I_{0}(\epsilon x)}\exp
(\epsilon \mathbf{x}\cdot \mathbf{\theta )},  \label{oo-eventdensity}
\end{equation}%
where $\mathbf{x}=(x\cos \phi ,x\sin \phi )$ and $\mathbf{\theta }=(\cos
\theta ,\sin \theta )$.
\begin{figure}[h]
\vspace{2ex}
\includegraphics[width=.4\textwidth]{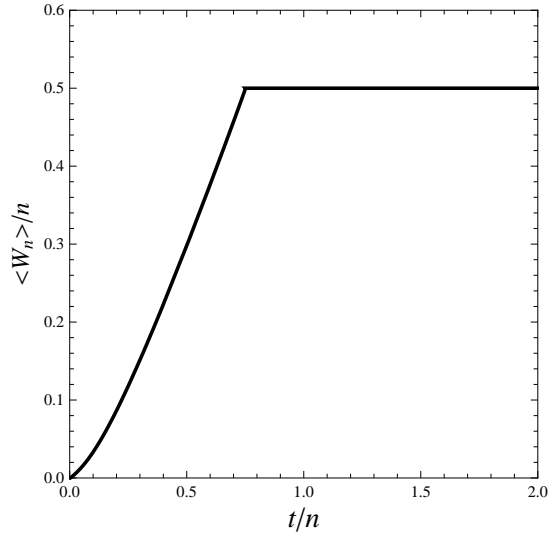}
\caption{Dependence of the active time per unit event $\langle W_{n}\rangle/n
$ for $n \gg 1$ on $t/n$ for the renewal
process model of correlated events designed to be analogous to the HMF
model. The figure shows the two-phase behavior described in the text. }
\end{figure}

\noindent The corresponding expression for $\psi_n(\epsilon; \tau, \theta)$, the
inverse Laplace transform of Eq.\hspace{1ex}(\ref{oo-eventdensity}) obtained via the saddle-point
approximation, is 
\begin{widetext}
\begin{equation}
 \psi(t;\tau,\theta)\simeq C\exp{(x^{2}-1/2)}\left(\frac{x^2-1/2}{t-\tau+{\bf
x}\cdot\theta}\right)^{1/2}\left(I_{0}\left[\frac{(x^{ 2 } -1/2)x}{t-\tau+{\bf x}\cdot
\theta}\right]\right)^{-1}, \label{1-eventdensity}
\end{equation}
\end{widetext}
where $C$ is a normalization constant. 
\begin{figure}[ht]
\includegraphics[width=.4\textwidth]{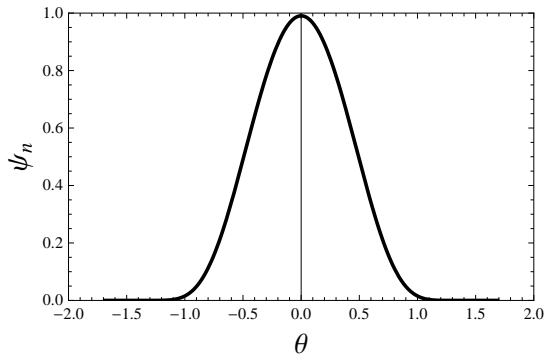}
\caption{Dependence of $\psi_{n}(t ;%
\protect\tau ,\protect\theta )$, $n\gg 1$, on the $\protect\theta $ when $%
t < t_{c}$. When $t >t_{c}$ the function $\psi_{n}(t
;%
\protect\tau ,\protect\theta )$ is $\protect\theta $-independent. }
\end{figure}
Following Refs.
\cite{ruffo2} \cite{ruffo3} we
evaluated the dependence of $\left\langle W_{n}\right\rangle /n$ on $%
\epsilon $ after solving numerically Eq.\hspace{1ex}(\ref{auxilliary1}). Subsequently
we used this in Eq. (\ref{caloric1}) to obtain the dependence of $t/n$ on $%
\epsilon $ (shown in Fig.\hspace{1ex}1, where the two-phase feature is evident). This
allowed us to determine the time dependence of the mean active time per
event, $\left\langle W_{n}\right\rangle /n$, shown in Fig.\hspace{1ex}2, where it is
observed that this quantity increases monotonically with $t$ until it
saturates at a value of $1/2$ at $t_{c}$ and remains constant thereafter.
Since $\left\langle W_{n}\right\rangle /n$ measures the average spread of
the tags $\theta _{i}$, $i=1,...,n$, we conclude that for short times $%
t<t_{c}$ this spread falls below its maximum whereas for larger times $%
t>t_{c}$ the maximum spread is always assured. In terms of the opinion poll
sets of $n$ samples taken inside time intervals $t<t_{c}$ suffer from
coverage bias but are as free of it as it is possible when the set of
samples are collected within time intervals $t>t_{c}$. This feature is
corroborated in Fig. 3 where we show the $\theta $-dependence of $\psi
_{n}(t;\tau ,\theta )$, $t<t_{c}$, as given by Ec.(\ref{1-eventdensity}).  When $t>t_{c}$ the
density is flat,
independent of $\theta $. Interestingly, sequences of $n$ events that take
place within time intervals $t<t_{c}$ are correlated while those for $t>t_{c}
$ are not. When $\left\langle W_{n}\right\rangle /n<1/2$ there is on average
no sufficient time for the pollster to move to other locations or to switch
to different population groups, there is a repetition, or ordering in the
set of samples. This generates a coverage bias. Similar arguments can be
elaborated in terms of the average idle time $\left\langle
T_{n}\right\rangle /n=1/(2\epsilon )$ that in the example of an opinion poll is
reflected by the presence of nonresponse bias. A measure of the correlations
induced for $t<t_{c}$ is given by the time derivative of $\left\langle
W_{n}\right\rangle /n$ (one of several response functions) as shown in Fig.
4. For $t>t_{c}$ the HMF model behaves effectively as an ideal gas, and, as
we can see from Figs. 2 to 4, the renewal process conforms to that of
independent events for this regime.

Thus, by construction our opinion poll renewal process acquires all the
properties of the HMF model, mainly its second order phase transition that
separates two different regimes. That is, for small $t/n$ strings of $n$
events cluster around a given value of the tag $\theta $ symptomatic of an
inefficient poll, but for larger $t/n$ the events display a uniform
dispersal of $\theta $ suggesting the proper working of the sampling
process. The clustering of the tag $\theta $ when $\epsilon >\epsilon _{c}$
is expressed by the Laplace transform variable $\epsilon $ as it measures
the width of $\widehat{\psi }_{n}(\epsilon ;\tau ,\theta )$. There are other
known interesting properties of the HMF model such as the occurrence of long
lived, or quasistationary, states for temperatures below the transition
temperature, when the system displays features of the high-temperature phase
uniform in $\theta $ \cite{ruffo2} \cite{ruffo3}. These states would
manifest also in the renewal process as sequences of active times $\langle W_n\rangle/n$ larger than
those shown
in Fig. 2 for some range of values $t < t_c$ close to $t_{c}$.
\begin{figure}[t]
\includegraphics[width=.4\textwidth]{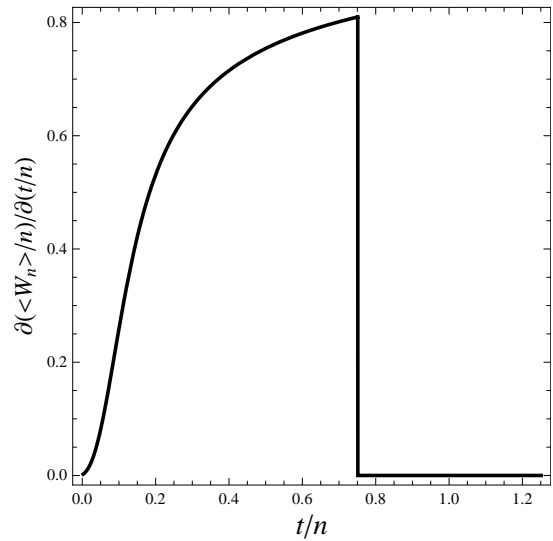}
\caption{Dependence of $\partial(\langle W_{n}\rangle/n)/\partial (t/n)
$, $t, n \gg 1$, on $t/n $. }
\end{figure}

\section{Summary and discussion}

We have made use of the statistical-mechanical interpretation of the basic
elements that constitute the theory of renewal processes. Our purpose for
recapturing this analogy is to facilitate the application of useful
techniques and approximations built up and tested through a large amount of
studies of thermal systems. Potentially these methodologies can have
important effects in the study of complex systems that originate outside
ordinary statistical-mechanical physical systems, in a variety of fields, in
ecology, economy, sociology, etc. where stochastic processes such as that
for the renewing of events often arise. The known parallels between renewal
processes and statistical mechanics are an indication of the general,
Laplace and Legendre transform structure of large deviations theory \cite%
{touchette1}. The saddle-point approximation is central to this theory where
a probability $P_{n}$ obeys the form $P_{n}\simeq \exp (-ns)$ for $n\gg 1$
with $s$ a positive quantity independent of $n$ named the rate function \cite%
{touchette1}. Clearly, the Massieu potentials $S_{\epsilon ,n}$ and $S_{t,n}$
in Eqs. (\ref{epsilonene4}) and (\ref{teene4}), respectively, when written
as $S_{\epsilon ,n}=-ns(\epsilon )$ and $S_{t,n}=-ns(\tau )$ comply with
this property. Thus we could describe renewal processes, familiar in the
probability theory domain, in this alternative language. Nevertheless,
because of our stated purposes we have used a statistical-mechanical
language. We have pointed out the equivalence with large deviation theory
when appropriate. As a difference from the present statistical study of
single sequences of correlated events, sets of renewal sequences dependent
on each other (when uncoupled the sequences are made of uncorrelated events)
have been analyzed with the use of multivariate distributions. For a recent application see Ref. \cite{Sumita}.

To exemplify the use of the parallelism between renewal processes and
statistical mechanics we devised a model renewal process with correlated
events that displays a phase transition as time progresses. When $n$ events
take place within a relatively short time interval their correlation is
evident, whereas for longer time intervals their statistical properties are
identical to those of independent events, and there is a sharp transition
between the two regimes. The renewal process ensemble structure facilitated
the description of this model that we chose to portray, amongst several
possible options, in terms of a polling process, and with characteristics
taken straightforwardly from a well-known particle or spin
statistical-mechanical model, the HMF model \cite{ruffo1}. There are examples of phase transitions
occurring along time evolution in deterministic (as opposed to stochastic) systems. See Ref.
\cite{Beck} and references therein.

We close by mentioning that correlated random walk processes on regular
lattices, as described with the help of the Fourier transform and generating
functions \cite{montroll1}, and for both discrete and continuous time
distributions, can be couched into a partition function language just as we
have shown here for correlated renewal processes. Due to the sign of the
integers used to locate the walker in lattice space the ensuing
statistical-mechanical formalism differs also from the canonical type,
basically by using the velocity instead of the kinetic energy as the primary
variable that describes interacting particles.

\begin{acknowledgments}
AR acknowledges an interesting conversation with
F. Baldovin. We are grateful for support from DGAPA-UNAM and CONACyT
(Mexican agencies).
AR acknowledges support from MEC (Spain).
\end{acknowledgments}

\bibliography{velazquezrobledo}

\begin{thebibliography}{15}
\bibitem [{not()}]{note1}%
  \BibitemOpen
  \href@noop {} {\bibinfo  {journal} {{The density $\chi _{n}(t)$ for the
  occurrence of $n$ events up to time $t$ is obtained form $\psi _{n}(t)$ by
  consideration of the probability $\Psi (t)=1-\int_{0}^{t}dt^{\prime }\ \psi
  (t^{\prime })\ =\int_{t}^{\infty }dt^{\prime }\ \psi (t^{\prime })$ that
  there is no renewal event in the time interval $(0,t)$. In Laplace space one
  obtains $\widehat{\chi }_{n}(\epsilon )=\epsilon ^{-1}(1-\widehat{\psi
  }(\epsilon ))\widehat{\psi }_{n}(\epsilon )$}}}\BibitemShut {NoStop}%
\end{thebibliography}

\end{document}